\newcommand{\pb}{\,\mbox{{\rm pb}}}
\def\GeV{\hbox{$\;\hbox{\rm GeV}$}}
\def\TeV{\hbox{$\;\hbox{\rm TeV}$}}
\newcommand{\Rp}{\mbox{$\not \hspace{-0.15cm} R_p$}}
\newcommand{\fb}{\mbox{{\rm ~fb}}}

\documentclass{ws-procs9x6}

\begin{document}

\title{Beyond the Standard Model at HERA: Status and Prospects}
%
\author{E. PEREZ
\footnote{\uppercase{T}alk given at the \uppercase{W}orkshop ``\uppercase{N}ew \uppercase{T}rends 
in \uppercase{H}\uppercase{E}\uppercase{R}\uppercase{A} \uppercase{P}hysics", \uppercase{R}ingberg
(\uppercase{G}ermany), \uppercase{O}ctober 2-7 2005.}}

\address{CE-Saclay, DSM/DAPNIA/Spp, F-91191 Gif-sur-Yvette \\
and DESY, Notkestrasse 85, D-22607 Hamburg. \\ 
E-mail: eperez@hep.saclay.cea.fr }

\maketitle

\abstracts{
An overview of experimental results on searches for new phenomena at HERA 
is presented. The complementarity with searches performed at other experiments
is discussed and the prospects for a discovery, using the full HERA data to
be delivered until mid-2007, are presented.}

\section{Introduction}

Although remarkably confirmed so far by low and high energy experiments,
the Standard Model (SM) of strong, weak and electromagnetic interactions
remains unsatisfactory. Many models of ``new physics" have been proposed
to address the questions which are unexplained by the SM. 
Experimentally, the observation of deviations with respect to the SM 
predictions is a key part of the existing and future high energy programmes.
At HERA, searches for new phenomena have been carried out using a luminosity
of up to $280 \pb^{-1}$. Besides an excess of atypical $W$-like events 
observed at H1 all measurements are so far in
good agreement with the SM expectation and constraints on models for new
physics have been obtained.

\section{Searches for new phenomena in inclusive DIS}

Neutral Current Deep Inelastic Scattering (NC DIS) is measured at HERA for
values of the photon virtuality $Q^2$ up to about $40000 \GeV^2$. Although
the precision is still statistically limited at the
highest $Q^2$, the good agreement of the measurements with the SM expectation
allows stringent constraints on new physics to be set~\cite{INCLUSIVE}. 
For example, a finite quark
radius would reduce the high $Q^2$ DIS cross section with respect to the SM predictions,
such that the current data rule out quark radii larger than $0.85 \cdot 10^{-18}$~m,
assuming that the electron is point-like.
Similarly, the effective scale of $eeqq$ contact interactions is constrained to be
larger than typically $5 \TeV$ - a similar sensitivity being achieved from the preliminary
Run II Drell-Yan data of the Tevatron experiments~\cite{D0_CONTACT_INTERACTIONS}.
The effective mass scale associated to the exchange of Kaluza-Klein gravitons
in models assuming additional large space dimensions is constrained to be larger
than about $0.8 \TeV$, slightly below the LEP and Tevatron bounds.

The longitudinal polarisation of the lepton beam in the HERA II data has been
exploited to measure the polarisation dependence of the Charged Current (CC)
DIS cross section~\cite{JOACHIM}. 
Although these data constrain in principle
the existence of a right-handed $W$ boson, the sensitivity is below the $W_R$
mass bounds obtained at the Tevatron. However these measurements confirm the left-handed nature
of the weak interaction in the $t$-channel.

\section{Model-dependent searches}

As HERA is not an annihilation machine, the pair-production of new heavy particles,
which could occur in $e^+ e^-$ or $pp$ collisions via their coupling to gauge bosons,
has a very low cross-section at HERA. Instead, searches for the {\it single}
production of new particles are performed at the H1 and ZEUS experiments. The
cross-section for such processes depends on the unkown coupling of the new
particle to the SM fields. Hence these searches do not provide {\it absolute}
constraints on the mass of new particles. Conversely, the observation of
a signal for the single production of a new particle would provide information
not only on its mass, but also on this unknown coupling. 
In the following example constraints obtained on 
leptoquarks, squarks in $R$-parity violating supersymmetry and excited
fermions are presented.

\subsection{Lepton-quark resonances: Leptoquarks}

\begin{figure}[htb]
\centerline{
\epsfxsize=3.1in\epsfbox{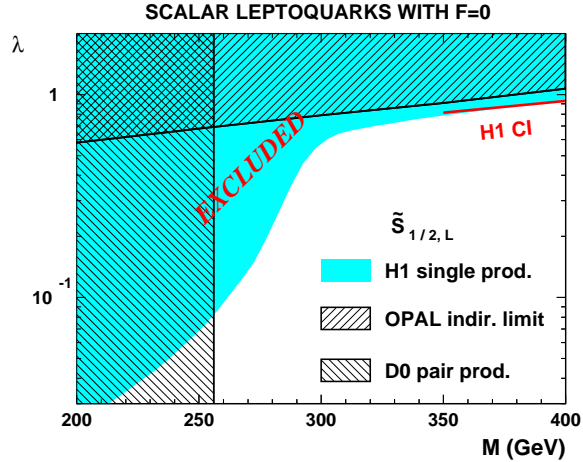}
}
\caption{Example mass-dependent upper bounds on the Yukawa coupling
$\lambda$
         of a first generation scalar leptoquark to the electron-quark pair.
         These are shown for a LQ of $F=0$ coupling to an $e^+$ and a $d$ quark, which 
         decays exclusively into $ed$.}
\label{fig:LQlim_lambda}
\end{figure}

An intriguing characteristic of the Standard Model is the observed symmetry
between the lepton and the quark sectors, which is manifest in the representation
of the fermion fields under the SM gauge groups, and in their replication over
three family generations. This could be a possible indication of a new symmetry
between the lepton and quark sectors, leading to ``lepto-quark" interactions.
Leptoquarks (LQs) are new scalar or vector colour-triplet bosons carrying
both a baryon and a lepton number.
Several types of LQs can be predicted, differing in their
quantum numbers. 
The interaction of the LQ with a lepton-quark pair is
parameterized by a coupling $\lambda$.
Depending on its quantum numbers a LQ couples to $eq$, $\nu q$ or both.
The branching ratio $\beta$ for a LQ to decay into an electron and a quark
can be fixed by model assumptions, or can be treated as a free parameter.
Decay modes other than $eq$ and $\nu q$ are usually neglected.

In electron-proton collisions, first generation
LQs might be singly produced via the fusion
of the incoming lepton with a quark coming from the proton.
The production cross-section roughly scales with $\lambda^2$ times the parton
density function of the relevant parton evaluated at $x_{Bj} = M^2_{LQ} / s$,
$\sqrt{s}$ being the centre-of-mass energy and $M_{LQ}$ the LQ mass.
Hence, $e^+ p$ ($e^- p$) collisions provide a larger sensitivity to LQs with
fermion number $F=0$ ($F=2$) 
since the $u$ and $d$ parton density is larger than that of antiquarks
at large $x_{Bj}$.
LQs might be observed as a resonant peak in the 
lepton-jet mass spectrum of NC or CC DIS events.
No such signal has been observed by the H1 and ZEUS experiments~\cite{HERALQ}.

\begin{figure}[tb]
\begin{picture}(10,100)(0,40)
\put(10.,-60.) {
\centerline{
\epsfxsize=2.8in\epsfbox{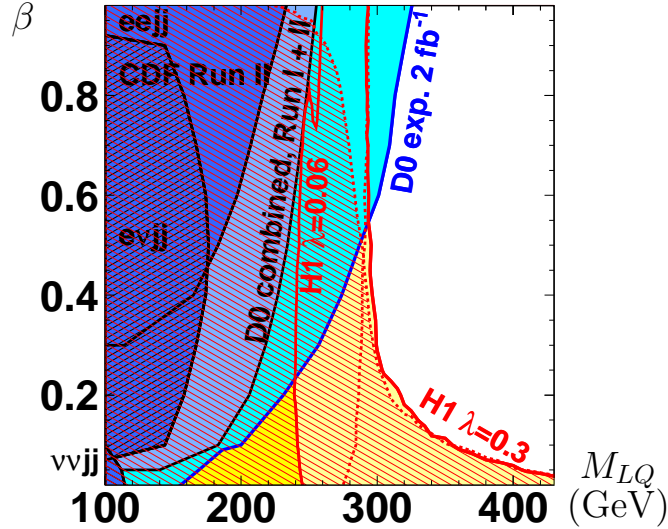}
}
}
\put(70,130){\Large {$\beta$}}
\put(285,-40){\Large {$M_{LQ}$} }
\put(280,-55){\Large {(GeV)}}
\end{picture}
\vspace*{3.5cm}
\caption{Example constraints on first generation scalar leptoquarks
         decaying exclusively into $eq$ and $\nu q$.
         For $\lambda=0.3$, the H1 constraints
         obtained from the $ej$ and $\nu j$ analyses are indicated by
         the dotted curves, and the combined bound is shown for two
         $\lambda$ values.
         The future sensitivity of the Tevatron
         for a luminosity of
         $2 \fb^{-1}$ is also shown.
        }
\label{fig:LQlim_beta}
\end{figure}

Existing constraints on a scalar leptoquark which decays solely
into an electron and a quark ($\beta = 1$) are summarised in
Fig.~\ref{fig:LQlim_lambda}.
For an electromagnetic strength of the coupling $\lambda$
($\lambda^2 / 4 \pi = \alpha_{em}$, i.e. $\lambda \simeq 0.3$), the HERA
experiments rule out LQ masses below $\sim 290 \GeV$.
%
%
Constraints derived from the search for pair produced
LQs at the
Tevatron do not depend on the coupling $\lambda$ and a lower
mass bound of $256 \GeV$ is set by the D0 experiment~\cite{D0_LQS}.
%
%
Example constraints on more general LQ models where $\beta \neq 1$ are shown in
Fig.~\ref{fig:LQlim_beta} assuming that the LQ decays exclusively into
$eq$ and $\nu  q$. Leptoquarks decaying with a large branching ratio
into $\nu  q$ are not easily probed at the Tevatron due to the
large background for final states containing only jets and missing transverse
momentum. In contrast, with a similarly good signal to background ratio 
for the $eq$ and $\nu q$ final states, HERA experiments can be 
sensitive on LQs which decay with a large branching
ratio into $\nu  q$, provided that the coupling $\lambda$ is not too small.


The current HERA bounds are more
stringent for $F=0$ LQs due to the much larger HERA~I luminosity in $e^+ p$
collisions (about $110 \pb^{-1}$ in $e^+ p$ and $15 \pb^{-1}$ in $e^- p$). 
Assuming the existence of a $F=0$ LQ in the HERA kinematic range,
and that its coupling to the $eq$ pair is equal to the HERA~I upper limit,
a significance of $\sim 4 \, \sigma$ could be reached in each experiment with an
$e^+ p$ luminosity of $350 \pb^{-1}$ and combining the H1 and ZEUS data could
allow the $5 \, \sigma$ threshold to be reached.
Assuming a LQ production cross-section
equal to half the existing limit the discovery potential is limited.
In contrast a much larger discovery potential remains
for $F=2$ leptoquarks,
which will be probed with the $e^- p$ data to be delivered
until mid-2007.

It should be noted that an $ep$ collider would be the ideal machine to study
a leptoquark signal. The fermion number of the leptoquark could easily be
determined by comparing the signal rates in $e^+ p$ and $e^- p$ collisions; the
polarisation of the lepton beam provides a handle to determine the chiral couplings
of the LQ; and the good signal to background ratio in the $\nu q$ channel allows
the LQ coupling to the neutrino to be studied.

\subsection{R-parity violating supersymmetry}

\begin{figure}[b]
\centerline{
\epsfxsize=2.9in\epsfbox{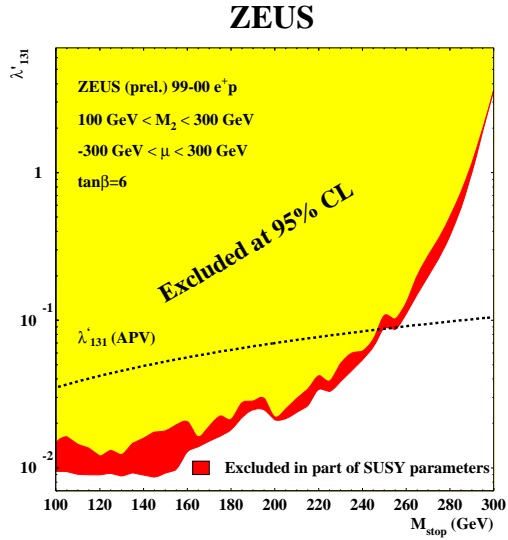}
}
\caption{Example mass-dependent constraints on the coupling of the stop to an $e^+ d$ pair.
         A scan of the SUSY parameter space has been performed. The light shaded domain
         is ruled out at $95 \%$ confidence level for any value of the parameters, which
         determine the dominant final states in which the signal could be observed.
         As can be seen from the narrowness of the dark band, the sensitivity
         of the analysis is nearly model-independent.}
\label{fig:SUSY}
\end{figure}

In supersymmetric (SUSY) models where the so-called $R$-parity ($R_p$) is not conserved, squarks
could be resonantly produced at HERA similarly to leptoquarks. In addition to the
``LQ-like" decays into $eq$ and possibly $\nu q$ the squarks also undergo decays into
gauginos (the supersymmetric partners of gauge bosons) and an exhaustive search
requires a large number of
final states to be analysed. This has been pioneered
by the H1 collaboration in~\cite{H1_SUSY} where the full HERA~I dataset has been
used to set constraints on supersymmetric models. 
A similar preliminary analysis looking for the SUSY partner of the top quark has
been performed by the ZEUS experiment. Example constraints are shown in Fig.~\ref{fig:SUSY}.
For a squark $R_p$-violating coupling of the electromagnetic strength, lower mass
bounds of $270-280 \GeV$ can be set. Within constrained SUSY models where a few
parameters determine the full Higgs and supersymmetric spectrum, stringent bounds were
derived on the squark mass from searches for Higgs, sfermions and gauginos at LEP. However, part of
the SUSY parameter space remains open for a discovery at HERA~II, for a
reasonably large $\Rp$ coupling~\cite{H1_SUSY}. The case of a light stop or sbottom is of high
interest for HERA~II since the bounds coming from Tevatron are less stringent than those
obtained assuming five degenerate squarks.
In particular, the sensitivity on the sbottom, which has a larger
production cross-section in $e^- p$ than in $e^+ p$ collisions 
($e^- u \rightarrow \tilde{b}$), will considerably
increase with the HERA~II $e^- p$ data.

In case the squarks are too heavy to be produced at HERA, the $t$-channel exchange of a selectron
or sneutrino could allow for single gaugino production. This process has been considered in
two classes of SUSY models, differing in the dominant decay mode of the produced 
gaugino~\cite{GAUGINO}.
In both cases the analyses slightly improve the previous bounds
if the relevant R-parity violating coupling is quite large.
These are the first SUSY constraints set at HERA which are independent of the squark sector.

\subsection{Fermion-boson resonances}

\begin{figure}[bt]
\centerline{
\epsfxsize=2.9in\epsfbox{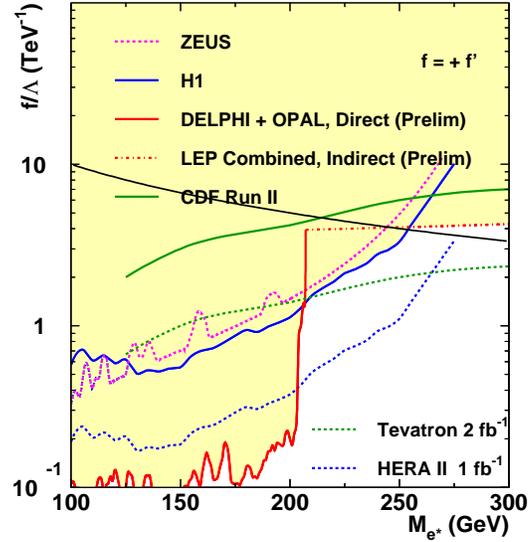}
}
\caption{Existing constraints on excited electron masses and
         couplings, assuming that $f=f'$. The decreasing curve
         shows the hyperbola $f / \Lambda = 1 / M_{e^*}$.
         The future HERA and Tevatron sensitivities are also shown
         as dotted curves.}
\label{fig:ESTAR}
\end{figure}

The observed replication of three fermion families motivates
the possibility of a yet unobserved new scale of matter.
An unambiguous signature for a new
scale of matter would be the
direct observation of excited states of fermions ($f^*$), via their decay
into a fermion and a gauge boson.
In the most commonly used model~\cite{HAGIWARA}, the interaction
of an $f^*$ with a fermion and a gauge boson is described by a magnetic
coupling proportional to $1 / \Lambda$
where $\Lambda$ is a new scale.
Proportionality constants $f$, $f'$ and $f_s$ result in different
couplings to $U(1)$, $SU(2)$ and $SU(3)$ gauge bosons.
Existing constraints on excited electrons are shown in
Fig.~\ref{fig:ESTAR}, under the assumption that $f=f'$.
Searches for pair produced $e^*$ at LEP
allowed to rule out
masses below about $103 \GeV$, independently of the value
of the coupling $f / \Lambda$. In contrast, searches for single
$e^*$ production at LEP, HERA~\cite{HERA_ESTAR} and Tevatron~\cite{CDF_ESTAR}
set mass bounds which depend
on $f / \Lambda$. 
The future HERA and Tevatron sensitivities, also depicted in
Fig.~\ref{fig:ESTAR}, show the discovery potential of HERA~II for
excited electrons.
The case of excited neutrinos is also very interesting for HERA~II,
since their production cross-section is larger
in $e^- p$ than in $e^+ p$ collisions by typically one order of magnitude.

\section{Searches for deviations from the SM in rare processes}

\label{section:rare}

\subsection{The ``isolated lepton events"}

Within the Standard Model, $W$ production at HERA has a cross-section of about $1 \pb$.
When the $W$ decays leptonically, the final state contains an isolated lepton,
missing transverse momentum, and a usually soft hadronic system. This process
has been measured using the HERA~I data~\cite{H1_ISOLATED_LEPTONS} in the ``electron"
($W \rightarrow e \nu_e$) and ``muon" ($W \rightarrow \mu \nu_{\mu}$) channels,
and a general agreement
with the SM prediction was observed. However, for large values of the transverse
momentum of the hadronic system, $P_T^X$, an excess of events was reported by
the H1 Collaboration~\cite{H1_ISOLATED_LEPTONS}. This excess was not confirmed by
a ZEUS analysis~\cite{ZEUS_TOP}, differing from the H1 analysis in terms of
background rejection\footnote{The non $W$ contribution to the expected background
was about $50 \%$ at large $P_T^X$ in the ZEUS analysis, while it amounts to 
$15 \%$ only in the H1 analysis.}.



\begin{figure}[bt]
  \centerline{
  \begin{tabular}{cc}
  \epsfxsize=2.3in\epsfbox{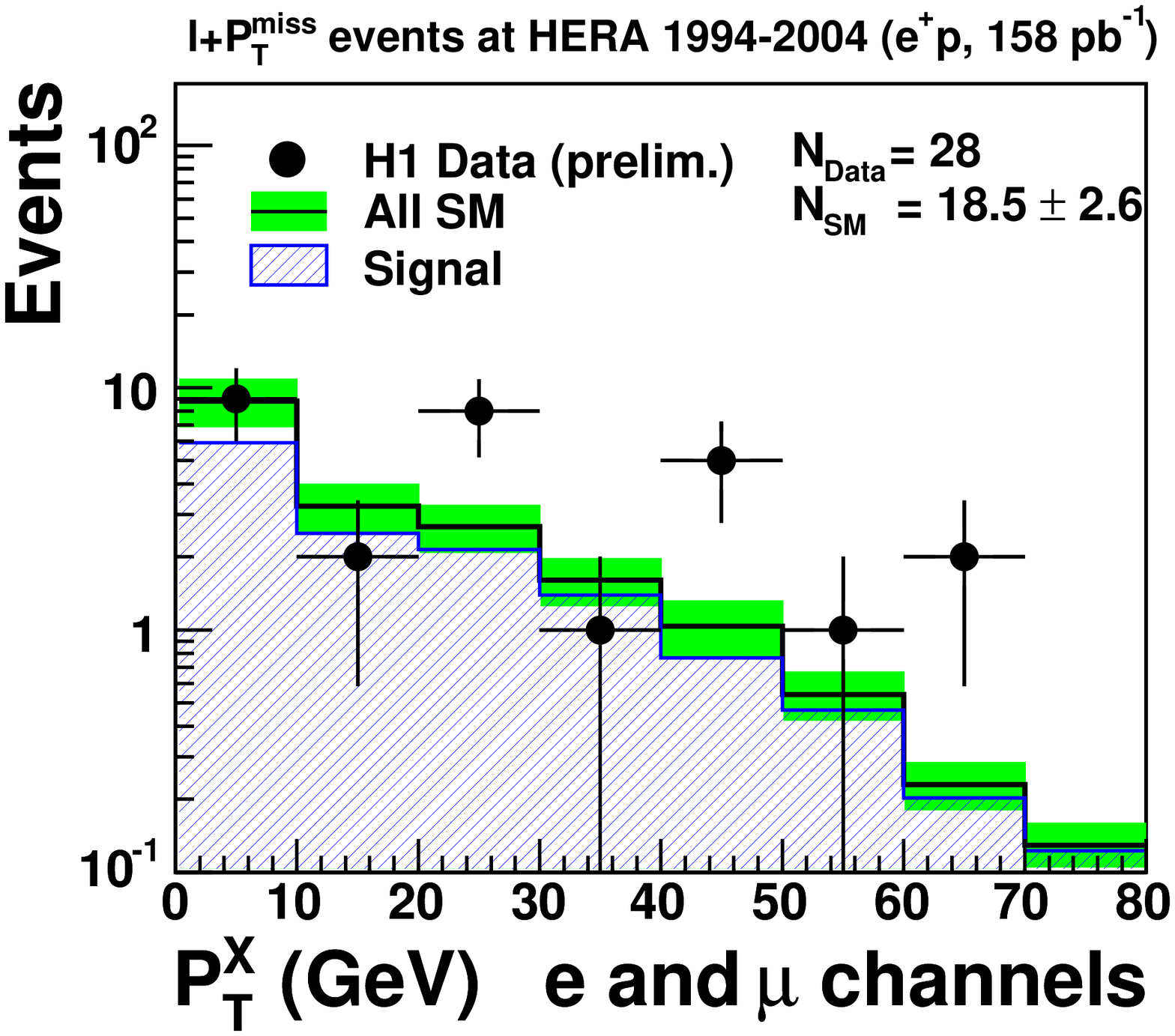} &
  \epsfxsize=2.3in\epsfbox{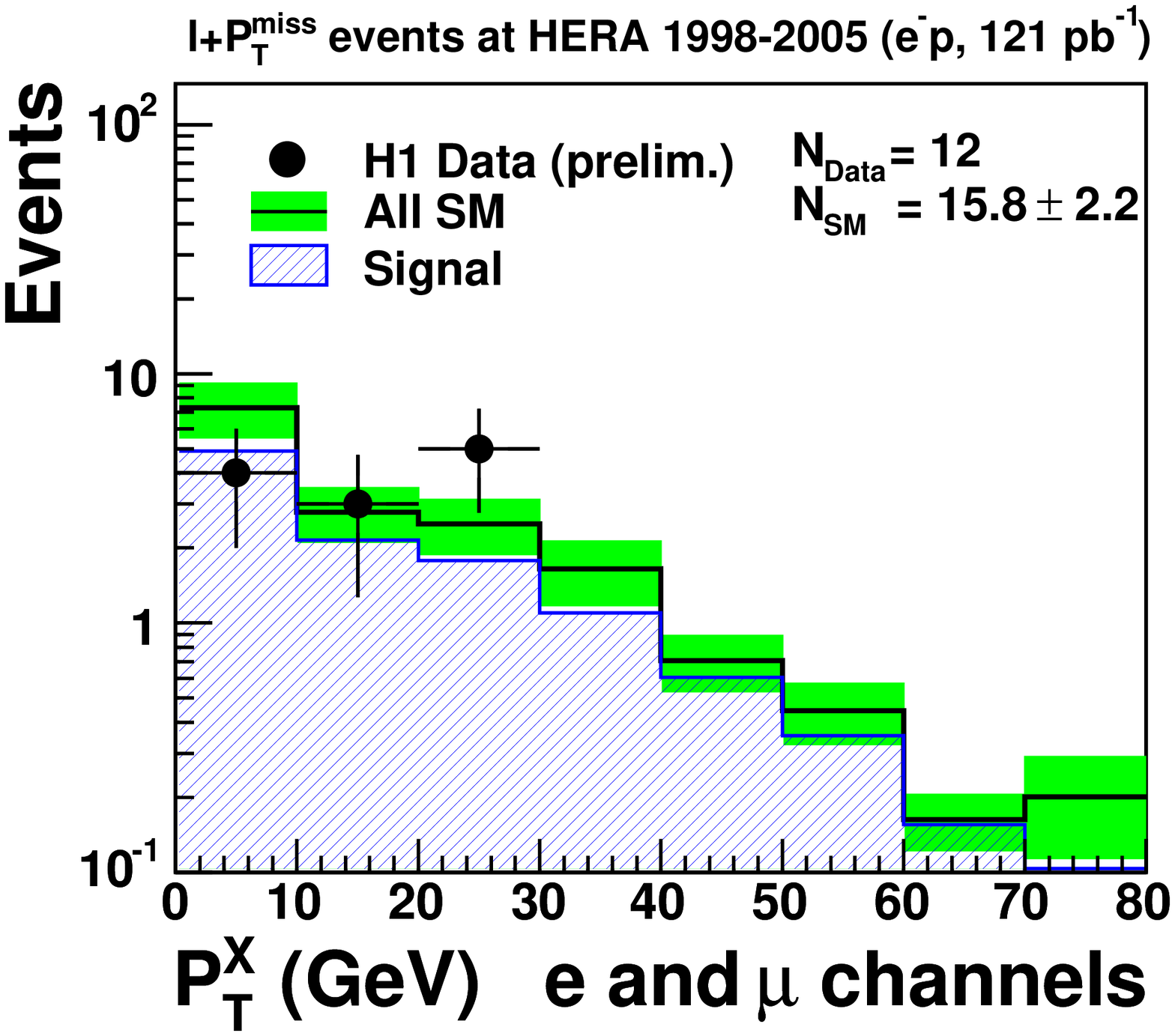}
  \end{tabular}
  }
  \caption{Distribution of the transverse momentum of the hadronic system
   $P_T^X$ in selected events recorded in (left) the $e^+ p$ data sample
   and (right) the $e^- p$ data sample.
   The hatched histogram shows the expectation
   from $W$ production while the total expectation is given by the open histogram.}
  \label{fig:separ}
\end{figure}


An abnormally large rate of high $P_T^X$ events is also observed by the 
H1 experiment~\cite{H1_ISOLATED_LEPTONS_UPDATE}
in the HERA~II data. Combining the $e$ and $\mu$ channels and the
HERA~I and HERA~II datasets,
which amount to a total luminosity of $279 \pb^{-1}$, $17$ events are observed
at $P_T^X > 25 \GeV$ for a SM expectation of $9.0 \pm 1.5$. Amongst the $6$ new
events observed in HERA~II, $5$ were recorded during the $e^+ p$ running
($53 \pb^{-1}$) and one
during the $e^- p$ running ($107 \pb^{-1}$).
Fig.~\ref{fig:separ} shows the observed $P_T^X$ distributions
separately for the $e^+ p$ and $e^- p$ datasets where HERA~I and HERA~II
data are combined, together with the corresponding SM expectations.
The observed and expected numbers of events are given in table~\ref{tab:separ}.
While the observation in the $e^- p$ data is consistent
with the SM expectation, $15$ events are observed at $P_T^X > 25 \GeV$ in the
$e^+ p$ data for an expectation of $4.6 \pm 0.8$ events. The probability that
the expected yield fluctuates to $15$ events or more corresponds to a
$3.4 \, \sigma$ fluctuation.

The ZEUS experiment has recently carried out a re-analysis of 
the $e$ channel using the 99-00 $e^+ p$
data, resulting in a larger purity in $W$ events.
The positron data taken at HERA~II have been analysed in the
same way, such that the total luminosity amounts to $106 \pb^{-1}$.
The results are also shown in table~\ref{tab:separ}.
At $P_T^X > 25 \GeV$ one event is observed in the data, in agreement with
the SM expectation
of $1.5 \pm 0.18$. 

Although the rate of events observed in the $e$ channel in the positron
data is larger in H1 than in ZEUS, both experiments are compatible
with each other within $2.5 \, \sigma$ for an average rate of about 4 events per $100 \pb^{-1}$
in this channel.
Assuming that such events are observed in the future H1 and ZEUS data at an average 
rate of about $7-8$ events 
per $100 \pb^{-1}$ combining the $e$ and $\mu$ channels, a significance of $4 \, \sigma$
could be reached\footnote{This estimate is obtained by scaling the SM backgrounds
of the H1 analysis in both the $e$ and $\mu$ channels.} from the combined H1 and ZEUS 
datasets by doubling the $e^+ p$ luminosity.
It should be noted that
new physics scenarios can be found which could explain that such events are observed
in $e^+ p$ collisions only. For example, in supersymmetry with two $R$-parity violating
couplings involving third generation fields, a top quark could be produced via 
$t$-channel sbottom 
exchange in $e^+ d$ collisions. Due to the large value of Bjorken $x$ needed to
produce a top quark in the final state, the corresponding process in $e^- p$
collisions would have a much lower cross-section.



\begin{table}[tbh]
\tbl{Summary of the H1 results of searches for events with
isolated electrons or muons and missing
transverse momentum for the $e^{+}p$ data
(158 pb$^{-1}$) and the $e^{-}p$ data (121 pb$^{-1}$). Data from HERA~I and
HERA~II are combined.
The number of events observed by ZEUS in the electron channel, in $106 \pb^{-1}$ of $e^+ p$ data,
is also shown.
The number of observed events at $P_T^X > 25 \GeV$ is compared to the SM prediction.
The quoted errors contain statistical and systematic uncertainties added in quadrature.
}
{\footnotesize
  \begin{tabular}{|c|c|c|c|c|}
    \hline
    \multicolumn{2}{|c|}{} &
    Electron &
    Muon &
    Combined \\
    \multicolumn{2}{|c|}{\large $P_T^X > 25 \GeV$} &
    obs./exp. &
    obs./exp. &
    obs./exp. \\
    \hline
    \hline
    \begin{tabular}{c}
    {H1} \\
    {Preliminary}
    \end{tabular}
    &
    \begin{tabular}{c}
    {\footnotesize 1998-2005 $e^{-} p$} \\
    {\footnotesize 121 pb$^{-1}$}
    \end{tabular}
    &
    {\footnotesize  2 /  2.4 $\pm$ 0.5 }&
    {\footnotesize  0 /  2.0 $\pm$ 0.3 }&
    {\footnotesize  2 /  4.4 $\pm$ 0.7 }\\
    \hline
    \hline
    \begin{tabular}{c}
    {H1} \\
    {Preliminary}
    \end{tabular}
    &
     \begin{tabular}{c}
    {\footnotesize 1994-2004 $e^{+} p$} \\
    {\footnotesize 158 pb$^{-1}$}
    \end{tabular}
    &
    {\footnotesize  9 /  2.3 $\pm$ 0.4 }&
    {\footnotesize  6 /  2.3 $\pm$ 0.4 }&
    {\footnotesize 15 /  4.6 $\pm$ 0.8 }\\
    \hline
    \hline
    \begin{tabular}{c}
    {ZEUS} \\
    {Preliminary}
    \end{tabular}
    &
     \begin{tabular}{c}
    {\footnotesize 1999-2004 $e^{+} p$} \\
    {\footnotesize  106 pb$^{-1}$}
    \end{tabular}
    &
    {\footnotesize 1 / 1.5 $\pm$ 0.18} & \multicolumn{2}{c}{} \\
    \cline{1-3}
  \end{tabular}\label{tab:separ}   } 
\end{table}


\subsection{Multi-lepton events}

If the events reported above were to be explained by some anomalous $W$ production
mechanism, an anomalous rate for $Z$-like events could also be observed.
Events with at least two electrons or muons in the final state have been looked for by
the H1 collaboration~\cite{H1_MULTILEPTONS}.
A slight excess of high mass multi-electron events was observed in the HERA~I
dataset. The analysis has been repeated using the HERA~II data and extended to
include other multi-lepton topologies~\cite{H1_MULTILEPTONS_HERAII}. 
With a total luminosity of $209 \pb^{-1}$
four events are observed with $\sum_{leptons} P_T > 100 \GeV$, three of which
being HERA~I $ee$ events. This is slightly above the SM prediction
of $0.81 \pm 0.14$.

\section{Conclusions}

Although some stringent bounds on new physics are set at LEP and
the Tevatron, HERA appears to be very well suited 
to search for new phenomena in some specific cases. 
In particular, searches for new physics at HERA rarely
suffer from huge SM backgrounds. Searches for leptoquarks, for the supersymmetric partners of the
top or bottom quarks, and for excited fermions might bring a discovery with the HERA~II data.
This holds in particular for new physics processes for which the cross-section
is larger in $e^- p$ than in $e^+ p$ collisions, since the HERA~I constraints are not too stringent
in such cases. With the future $e^+ p$ data to be delivered until mid-2007, the excess of atypical 
$W$-like events observed at H1, which corresponds to a $3.4 \, \sigma$ deviation,
appears to be the best chance for a discovery at HERA~II.


\begin{thebibliography}{0}
%
\bibitem{INCLUSIVE} H1 Collab., C. Adloff et al., Phys. Lett. {\bf B568} (2003) 35;
ZEUS Collab., S.~Chekanov et al., Phys. Lett. {\bf B591} (2004) 23. 
%
\bibitem{D0_CONTACT_INTERACTIONS} D0 Collab., D0 Note 4552-CONF.
%
\bibitem{JOACHIM} J. Meyer, these proceedings.
%
\bibitem{HERALQ} H1 Collab., Phys. Lett. {\bf B629} (2005) 9;
ZEUS Collab., S. Chekanov et al., Phys. Rev. {\bf D68} (2003) 052004. 
%
\bibitem{D0_LQS} D0 Collab., V.M. Abazov et al., Phys. Rev. {\bf D71} (2005) 071104.
%
\bibitem{H1_SUSY} H1 Collab., A. Aktas et al., Eur. Phys. J. {\bf C36} (2004) 425.
%
\bibitem{GAUGINO} H1 Collab., A. Aktas et al., Phys. Lett. {\bf B616} (2005) 31;
ZEUS Collab., contributed paper to EPS'05, abstract \#329; {\it idem}, contributed paper
to EPS'05, abstract \#330.
%
\bibitem{HAGIWARA} K.~Hagiwara, D.~Zeppenfeld and S.~Komamiya, Zeit. fur Phys. {\bf C29} (1985) 115;
F.~Boudjema, A.~Djouadi and J.L.~Kneur, Zeit. fur Phys. {\bf C57} (1993) 425.
%
\bibitem{HERA_ESTAR} H1 Collab., C. Adloff et al., Phys. Lett. {\bf B548} (2002) 35;
ZEUS Collab., S.~Chekanov et al., Phys. Lett. {\bf B549} (2002) 32. 
%
\bibitem{CDF_ESTAR} CDF Collab., D. Acosta et al., Phys. Rev. Lett. {\bf 94} (2005) 101802. 
%
\bibitem{H1_ISOLATED_LEPTONS} H1 Collab., V. Andreev et al., Phys. Lett. {\bf B561} (2003) 241.
%
\bibitem{ZEUS_TOP} ZEUS Collab., S. Chekanov et al., Phys. Lett. {\bf B559} (2003) 153.
%
\bibitem{H1_ISOLATED_LEPTONS_UPDATE} H1 Collab., document prepared for the Nov. 2005 DESY-PRC meeting,
available at http://www-h1.desy.de/publications/H1preliminary.short\_list.html.
%
\bibitem{H1_MULTILEPTONS} H1 Collab., A. Aktas et al., Eur. Phys. J. {\bf C31} (2003) 17; 
H1 Collab., A. Aktas et al., Phys. Lett. {\bf B583} (2004) 28.
%
\bibitem{H1_MULTILEPTONS_HERAII} H1 Collab., contributed to EPS'05, abstract \#635.
\end{thebibliography}
\end{document}